\documentclass[%
superscriptaddress,
 amsmath,amssymb,
 aps,
pra,twocolumn,
]{revtex4-2}

\usepackage{graphicx}
\usepackage{dcolumn}
\usepackage{bm}
 \usepackage{multirow}
\usepackage{hyperref}

\usepackage{xcolor}

\begin{document}


\title{GAN Decoder on a Quantum Toric Code for Noise-Robust Quantum Teleportation}

\author{Jiaxin Li}
\affiliation{Information Technology Department, Qingdao Vocational and Technical College of Hotel Management, Qingdao 266100, China}
\affiliation{College of Information Science and Engineering, Ocean University of China, Qingdao 266100, China}
\affiliation{Dipartimento di Ingegneria, Universit\`{a} di Palermo, Viale delle Scienze, 90128 Palermo, Italy}

\author{Zhimin Wang}%
\affiliation{College of Information Science and Engineering, Ocean University of China, Qingdao 266100, China}

\author{Alberto Ferrara}
\affiliation{Dipartimento di Ingegneria, Universit\`{a} di Palermo, Viale delle Scienze, 90128 Palermo, Italy}

\author{Yongjian Gu}
\email{yjgu@ouc.edu.cn}
\affiliation{College of Information Science and Engineering, Ocean University of China, Qingdao 266100, China}

\author{Rosario Lo Franco}
\email{rosario.lofranco@unipa.it}
\affiliation{Dipartimento di Ingegneria, Universit\`{a} di Palermo, Viale delle Scienze, 90128 Palermo, Italy}%


\begin{abstract}

We propose a generative adversarial network (GAN)-based decoder for quantum topological codes and apply it to enhance a quantum teleportation protocol under depolarizing noise. By constructing and training the GAN’s generator and discriminator networks using eigenvalue datasets from the code, we obtain a decoder with a significantly improved decoding pseudo-threshold. Simulation results show that our GAN decoder achieves a pseudo-threshold of approximately $p=0.2108$, estimated from the crossing point of logical error rate curves for code distances $d=3$ and $d=5$, nearly double that of a classical decoder under the same conditions ($p\approx 0.1099 $). Moreover, at the same target logical error rate, the GAN decoder consistently achieves higher logical fidelity compared to the classical decoder. When applied to quantum teleportation, the protocol optimized using our decoder demonstrates enhanced fidelity across noise regimes. Specifically, for code distance $d=3$, fidelity improves within the depolarizing noise threshold range $P<0.06503$; for $d=5$, the range extends to $P<0.07512$. Moreover, with appropriate training, our GAN decoder can generalize to other error models. This work positions GANs as powerful tools for decoding in topological quantum error correction, offering a flexible and noise-resilient framework for fault-tolerant quantum information processing. 
 
\end{abstract}

\maketitle

\section{\label{sec:level1}Introduction}

Quantum error correction codes (QECCs) \cite{1,2,3} are an important tool for solving the problem of noise interference and improving quantum fidelity. Quantum topological codes are considered among the most promising QECCs due to their high experimental feasibility and good performance \cite{4,5,6}. Quantum topological codes are highly degenerate \cite{7}, i.e., there are multiple strategies for correcting one error syndrome, thus a decoder is necessary to select a optimal error correction strategy. Finding high-fidelity and efficient decoders for quantum topological codes has been the subject of many studies. For example, Minimum weight perfect matching (MWPM) \cite{8,9} determines the optimal decoding strategy by finding the shortest error correction path for the syndrome. However, for the fidelity performance of quantum topological codes, decoders that lead to higher fidelity need to be further investigated, and deep learning has a good potential for development in this area \cite{10}. For example, a surface code decoder based on a recurrent neural network (RNN) has been developed  \cite{10a}. Also, a toric code decoder based on reinforcement learning (RL), achieving performance comparable to MWPM, has been proposed \cite{10b}. In addition, RL has been used to develop a decoder for continuous variables in the bosonic code space \cite{10c}.

Generative Adversarial Network (GAN) is a strong tool for deep learning that has shown great potential and wide range of applications in several fields. A typical GAN consists of a generator and a discriminator \cite{11}. The goal of the generator is to generate data that can fool the discriminator. The goal of the discriminator is to be able to accurately distinguish between true and false data. The generator and the discriminator go through a learning process of adversarial game and finally reach a Nash equilibrium point, which can better accomplish their goals. Currently, GAN is widely used in image generation, image resolution enhancement, data enhancement, speech generation and other fields \cite{13,15,16,17,18,19,20,21}. The error correction strategy of a topological code can be expressed as a binary matrix of eigenvalues. Generative adversarial networks learn these matrices so that the generator directly generates the error correction strategies.

\begin{figure*}
\includegraphics[width=18cm]{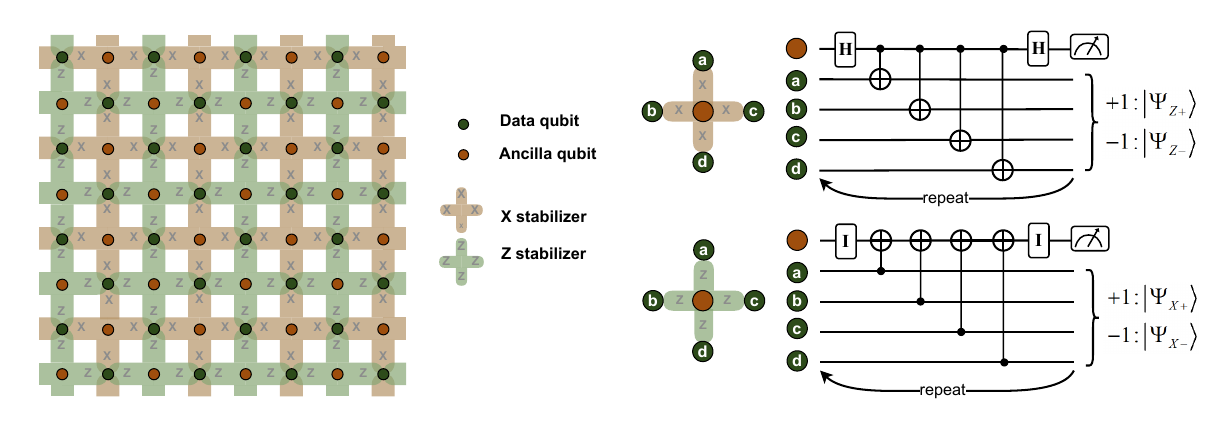}
\caption{\label{f1} The basic structure of toric codes. X stabilizers are used to detect Z errors, while Z stabilizers are used to detect X errors. The measurement circuits are shown on the right. H gates, I gates, and CNOT gates are used. The ancilla qubits are measured to obtain the eigenvalue. For a given stabilizer, the eigenvalue is +1 if the number of errors on the neighboring data qubits is even, and -1 if it is odd.}
\end{figure*}

Quantum communication is an important research direction in the field of quantum information science. It realizes the interaction of information under the rules of quantum communication protocols. Quantum teleportation \cite{22,23,24,24a,24b,24c} is a typical protocol in quantum communication, which has made new progress in both theory and experiment \cite{25,26,27,28}. In recent years, the achievable communication distance of quantum teleportation has been continuously upgraded. For example, in 2012, in quantum teleportation experiments using optical fibers and ground-based free-space channels, the transmission distance could reach about 100 kilometers \cite{29}. In 2017, based on satellite platforms and space-based connections, the communication distance of quantum teleportation was extended to 1,400 kilometers \cite{30}. In 2022, Hermans et al. used three optical connection points based on solid-state qubits to realize long-distance quantum teleportation between non-adjacent nodes in a quantum network \cite{31}.  However, due to the fragility of quantum states, effective quantum error correction and decoding methods are essential for improving the reliability of quantum teleportation \cite{31x,31y,31a,31b,31c,31d,namiki2016}. Here, we investigate decoding under a toric code and apply it to quantum teleportation to enhance its robustness under depolarizing noise.

In particular, we propose a novel GAN decoder for the quantum toric code. The primary aim of this work is to develop a decoder that achieves higher logical fidelity compared to some existing decoding methods. Furthermore, we demonstrate the application of the toric code with the GAN decoder in a quantum teleportation protocol, illustrating its potential to enable more noise-robust and high-fidelity quantum teleportation.  In this GAN decoder, two neural networks are used as generators and discriminators respectively. The trained generator generates error correction strategies with high success rates. This decoder has a higher pseudo-threshold and fidelity than MWPM. Specifically, below the threshold, toric codes have a higher error correction success rate at the same code distance, and larger code distances provide better decoding performance under the same noise rate. Here we apply the proposed GAN decoder on a quantum toric code to the teleportation protocol, but it can be extended to other quantum information protocols.

The outline of this paper is as follows. In Section \ref{sec:level2}, we give a brief background introduction to GAN, quantum teleportation, and topological toric codes. Section \ref{sec:level3} introduces a detailed description of the algorithm model structure and basic principles. Section \ref{sec:level4} gives the simulation experimental results of the model and comparative analysis with related works. Finally, in Section \ref{sec:level5} we discuss results and prospects of this work.

\section{\label{sec:level2}Theoretical Background}

\subsection{Toric code}
Quantum toric codes are a class of quantum topological codes with a spatial property that is preserved under continuous deformation. The idea behind the toric code is to encode the information with "homological degrees of freedom" \cite{32,33}, every bond has a spin-$\frac{1}{2}$ degree of freedom. As shown in Figure \ref{f1}, the basic structure of toric codes is a square lattice with periodic boundaries embedded in a closed surface.

The quantum information in toric codes is distributed across multiple physical qubits on a lattice with code distance d. Larger code distances achieve better fidelity below the threshold. The core of the toric code is detecting errors by measuring operators called stabilizers. These stabilizers are divided into X stabilizers and Z stabilizers, which detect bit-flip errors and phase-flip errors, respectively. By periodically measuring these stabilizers, a set of eigenvalues, called syndromes, is obtained, enabling the detection and localization of errors.

For a toric code with a code distance of $d$, the size of the lattice $C$ is $d\times d$, and it has three main features: $f$ faces, $e$ edges, and $v$ vertices. A dual lattice ${C^*}$ is constructed for the toric code.  The idea is to map $f$ to the dual vertices $v^*$, $e$ to the dual edges $e^*$, and $v$ to the dual faces $f^*$. Here $v \in C \Leftrightarrow {f^*} \in {C^*}$, $f \in C \Leftrightarrow {v^*} \in {C^*}$, $e \in C \Leftrightarrow {e^*} \in {C^*}$. The data qubits of the toric code are located on $e$ and $e^*$. The operators at $v $ (${f^*} $) are the Z stabilizers, and the operators at $f $ (${v^*} $) are the X stabilizer, denoted as 
\begin{equation}
\begin{aligned}
{Z_v} = \prod\limits_{e|v \in {\partial _1}e} {{Z_e}} ,  {X_f} = \prod\limits_{e \in {\partial _2}f} {{X_e}}, 
\end{aligned}
\end{equation}
where ${\partial _1}e$ are the vertices of $\forall e$, ${\partial _2}f$ are the edges of $\forall f$. Vertices and faces anti-commute with each other, and generate the stabilizers of toric codes. This anti-commutation relation ensures that bit-flip and phase-flip errors are independently detectable and correctable. It also guarantees that the stabilizer group forms a valid quantum code space, which encodes logical qubits in topologically protected entangled states. 

Denote the number of vertices, edges and faces of the lattice as $V$, $E$ and $F$, respectively. The toric code has $V+F-2$ independent generators. It follows from the stabilizer theory \cite{4} that the number of encoded qubits is 
\begin{equation}
\begin{aligned}
k = E - \left( {V + F - 2} \right) = 2 - \chi  = 2g,
\end{aligned}
\end{equation}
where $\chi$ is the Euler characteristic, $g$ is the number of closed surface loops. A toric code with a code distance of $d$ has $V=F=d^2$, $2d^2$ qubits and $2d^2-2$ stabilizers.

\begin{figure}
	\includegraphics[width=8.5cm]{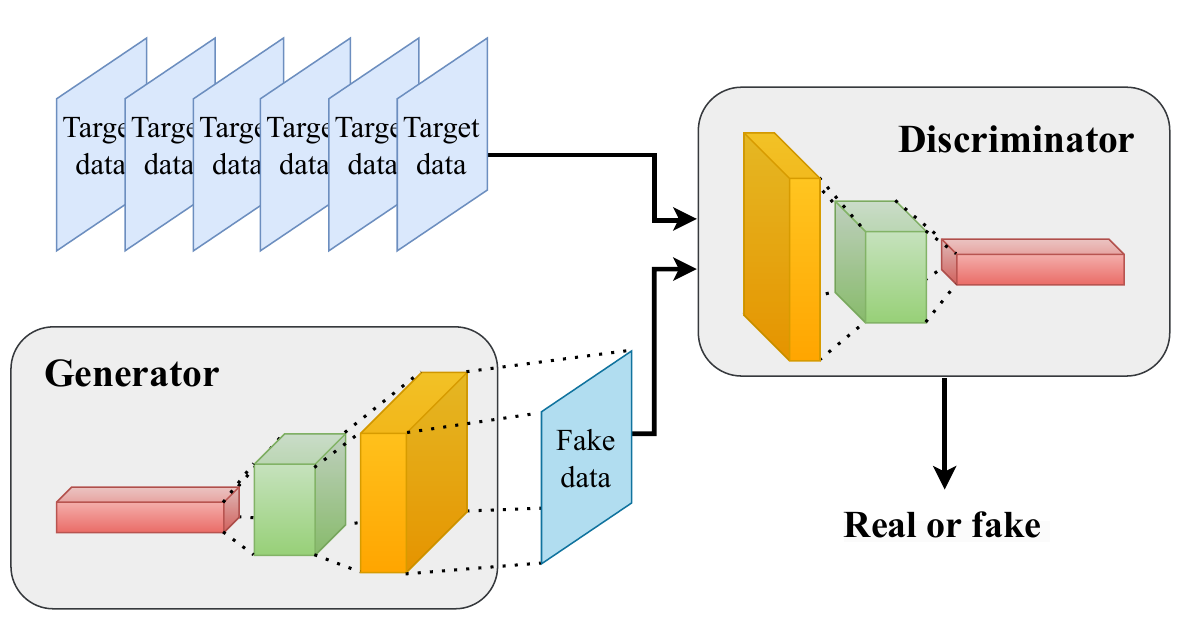}
	\caption{\label{f2} The structure of a GAN involves two components: the generator and the discriminator. The generator aims to produce fake data that approximates the target data, while the discriminator aims to distinguish between target and fake data. Through adversarial training, both the generator and discriminator are iteratively refined, leading to enhanced performance. The yellow, green, and red blocks represent different network layers of the GAN.}
\end{figure}
\subsection{Generative adversarial networks}
For a GAN, the generator and discriminator are typically provided with a deep learning network, such as a neural network or a convolutional network, which adjusts parameters in an attempt to find an optimal output. As shown in Figure \ref{f2}, the generator takes an input and transforms it through a series of neural network layers into fake data that resembles the training data. The discriminator receives a set of data samples as input, including real data (from the training set) and fake data generated by the generator. The task of the discriminator is to distinguish whether the input data is real or generated. It outputs a scalar value indicating the probability that the input data is real. During each training period, the parameters of generator are initially fixed while the discriminator is trained; then the parameters of the discriminator are fixed while the generator is trained. This process is repeated until the generated data resembles real data to the point that the discriminator can no longer effectively distinguish them.

\subsection{Quantum teleportation}
\begin{figure}
\includegraphics[width=9cm]{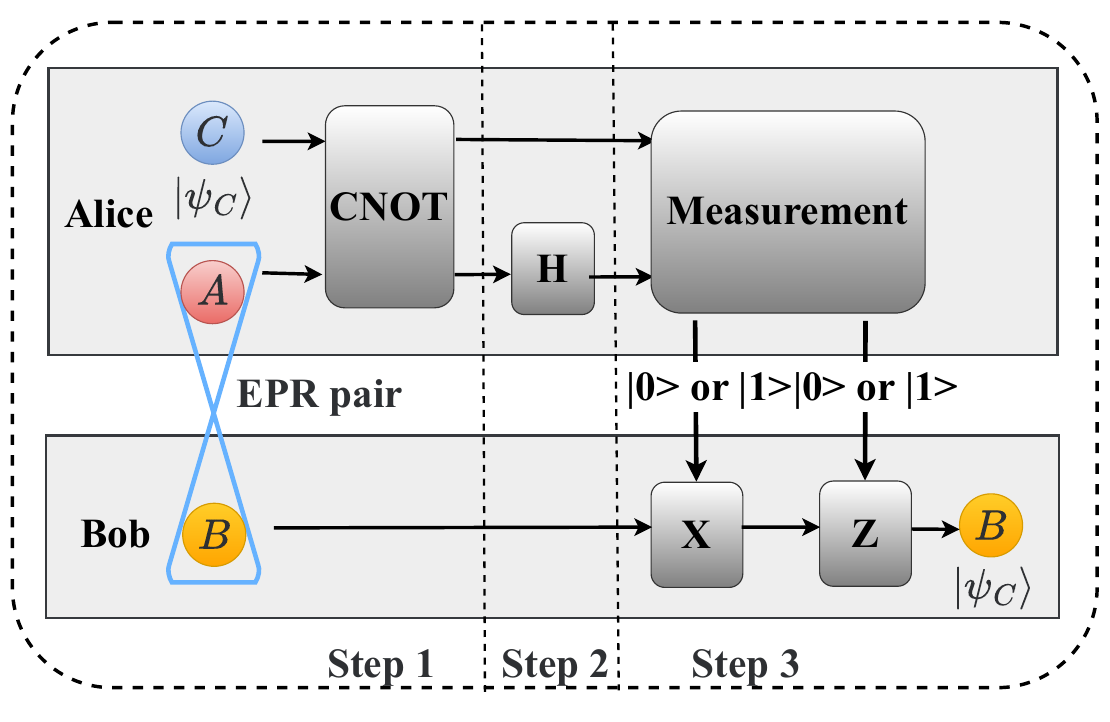}
\caption{\label{f3} Quantum teleportation. Particles $A$ and $B$ are the two particles of the EPR pair. The state of particle $C$, denoted as $\left| \psi_{C}  \right\rangle$, is to be teleported. Alice holds particles $A$ and $C$. Bob holds particle $B$. The gray rectangles represent quantum operations.}
\end{figure}

The quantum teleportation algorithm is a quantum communication method used to transfer quantum states. Suppose Alice wants to send the state $\left| \psi_{C}  \right\rangle = \alpha \left| 0 \right\rangle  + \beta \left| 1 \right\rangle $ of particle $C$ to Bob. First, Alice and Bob prepare a pair of EPR (Einstein–Podolsky–Rosen) particles. As shown in Figure \ref{f3}, particles $A$ and $B$ form an entangled state $\left| {{\beta _{00}}} \right\rangle  = {1 \mathord{\left/
 {\vphantom {1 {\sqrt 2 }}} \right.
 \kern-\nulldelimiterspace} {\sqrt 2 }}\left( {\left| {00} \right\rangle  + \left| {11} \right\rangle } \right)$. Alice holds particles $A$ and $C$, while Bob holds particle $B$. The quantum teleportation algorithm transmits $\left| \psi_{C}  \right\rangle $ to particle b through the following three steps:

1) Alice sets $\left| \psi_{C}  \right\rangle $ as the control qubit and $A$ as the target qubit to perform a CNOT gate. This operation entangles particle  $C$ with particles $A$ and $B$, forming a three-particle entangled system.

2) Alice applies a Hadamard gate to  $A$ and measures both  $A$ and  $C$, obtaining one of the four possible measurement results: $\left| 00 \right\rangle $, $\left| 01  \right\rangle $, $\left| 10  \right\rangle $, or $\left| 11  \right\rangle $. The outcome is sent to Bob  through a classical communication channel.

3) After the measurement, the probability information of  $\left| \psi_{C}  \right\rangle $ is transferred to $B$. Based on the measurement result, Bob performs no operation or applies one of the gates X, Z, or Y to $B$ to obtain the state  $\left| \psi_{C}  \right\rangle $.

\section{\label{sec:level3}Methods}
\subsection{Error correction}
Assume  $E_p = {X_c}{Z_{{c^*}}}$ represents a Pauli error, where $c$ denotes a chain (subset) of edges on the primal lattice indicating bit-flip errors, and $c^*$ denotes a chain of edges on the dual lattice indicating phase-flip errors. $X_c$ is the product of Pauli-X operators acting on the edges in $c$, and $Z_{c^*}$ is the product of Pauli-Z operators acting on the edges in $c^*$. The joint probability distribution is given by $p = \left\{ {{p_{c,{c^*}}}} \right\}$.
 The first step in error correction is to measure the stabilizers of the toric code and obtain the syndrome.  $Z_v$ and $X_f$ are used to detect the X error and Z error of the qubits, respectively. If errors occur, the eigenvalues of corresponding stabilizers change from $+1$ to 
$-1$, which means a pair of $-1$ defects are generated on the vertices of the lattice. There are ${d^2} - 1$ independent operators on a $d \times d$ lattice , thus a single $-1$ eigenvalue with all other $+1$ is impossible. The probability to obtain a given syndrome is 
\begin{equation}
\begin{aligned}
P\left( {\partial c,{\partial ^*}{c^*}} \right) = \sum\limits_{e \in {X_c}} {\sum\limits_{{e^*} \in {Z_{{c^*}}}} {P\left( {c + e,{c^*} + {e^*}} \right)} } .
\end{aligned}
\end{equation}

The syndrome is degenerate. For error correction, we need to select a set of correction qubits to eliminate the defects in the syndrome in pairs. If the correction qubits and error qubits form a trivial loop, the logical state of the toric code will return to its original state. However, if they form a nontrivial loop around the lattice, the error correction will fail. Although the nontrivial loop eliminates defects, it alters the original state of the toric code. The MWPM decoder selects the chain with the minimum amount of correction qubits. The success probability of this strategy is
\begin{equation}
\begin{aligned}
{P_{succ}} = \sum\limits_{\partial c,{\partial ^*}{c^*}} {\mathop {\max }\limits_{\left( {e,{e^*}} \right) \in E} P\left( {c + e,c + {e^*}} \right)}.
\end{aligned}
\end{equation}

\begin{figure}
\includegraphics[width=9cm]{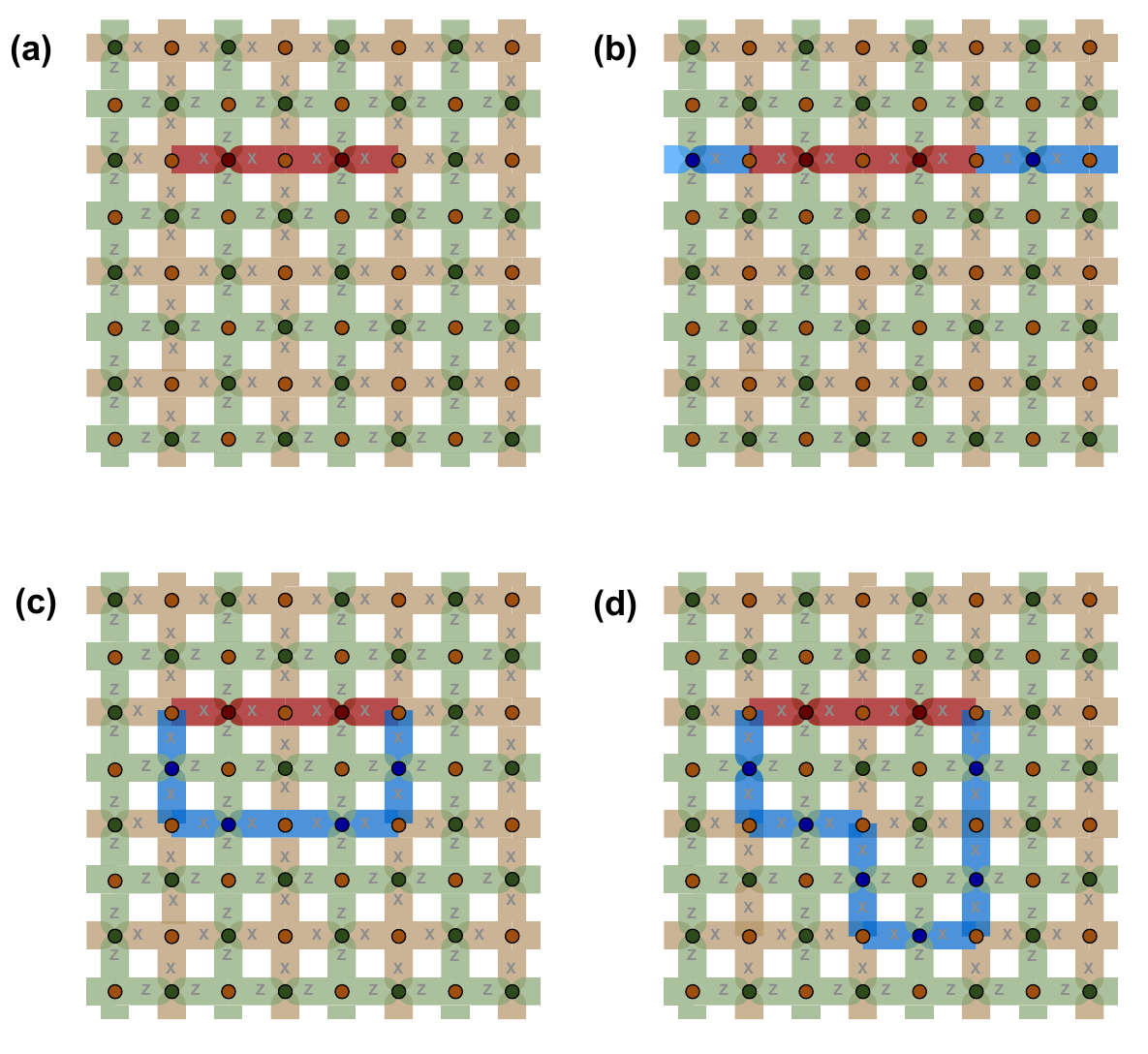}
\caption{\label{f4}Error correction of a quantum toric code. Red data qubits indicate that an error has occurred. Blue data qubits mean they are selected as an error correction qubit. (a) A toric code with error data qubits. The eigenvalues of the stabilizers around the error qubits becomes $-1$. (b) Failed quantum toric code error correction. A error correction chain and the error chain form a nontrivial loop. (c) Successful quantum toric code error correction. A error correction chain and the error chain form a trivial loop. (d) Another successful quantum toric code error correction. This strategy forms a trivial loop, but not the optimal. }
\end{figure}

In the limit of large lattices,  if the noise is below a critical threshold, ${P_{succ}} \to 1$. For the  incoherent noise model with $d \to \infty $, by mapping to the random bond Ising moldel, there is a critical threshold ${P_c} \approx 0.11$. This strategy certainly forms trivial loops when below this critical threshold. Figure \ref{f4} shows the error correction of a quantum toric code. In the figure, we intuitively illustrate the relationship between toric code qubits and stabilizers, and give examples of successful and failed error corrections for a defect caused by a string of error qubits.

For a string of errors, we need to find an error correction strategy. Since the error correction in toric codes is degenerate, there is not just one right correction scheme for a given set of error syndromes. This differs from classical non-degenerate error correction codes, where each error typically has a unique correction strategy. In toric codes, error correction is successful as long as the error qubits and correction qubits form a trivial loop. Conversely, forming a non-trivial loop means error correction has failed. Thus, the goal of our decoder is to generate a correction path such that the qubits on this path and the error qubits form a trivial loop.

\begin{figure*}
\includegraphics[width=11cm]{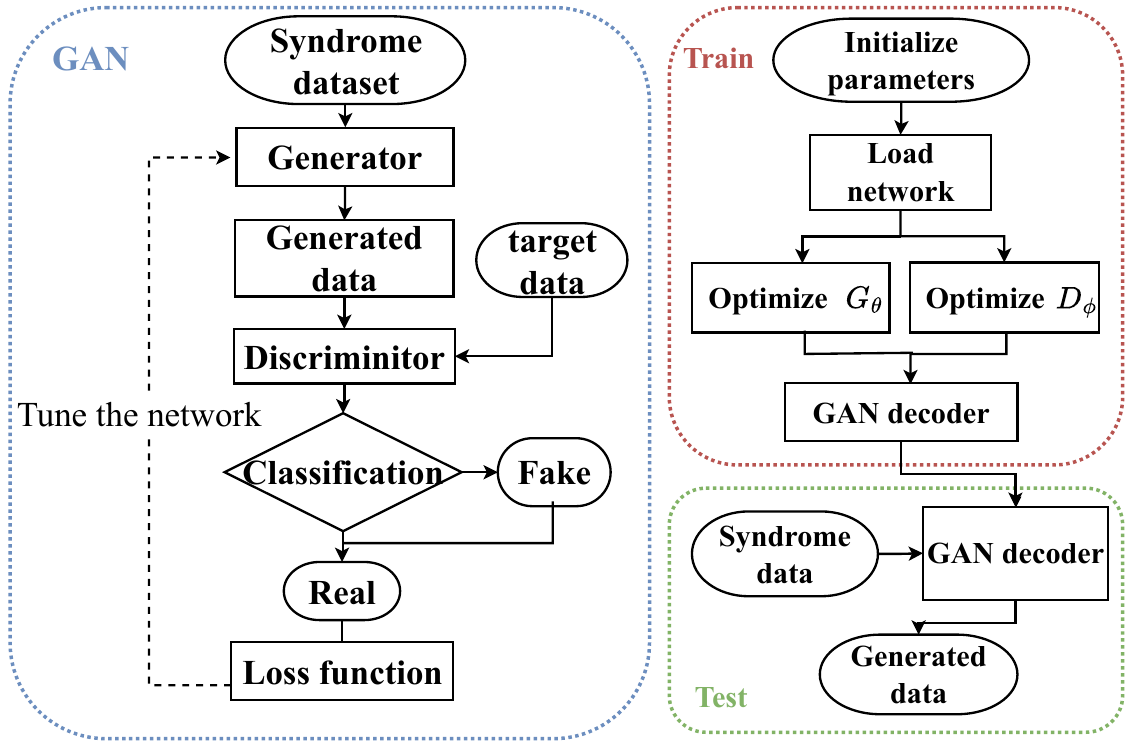}
\caption{\label{f5} The training framework for the GAN decoder. The blue dotted box illustrates the training process of the GAN, the red dotted box represents the parameter training process of the decoder, and the green dotted box depicts the testing process of the decoder.}
\end{figure*}

The standard MWPM is a global decoding algorithm. It works by pairing syndrome defects and selecting the matching path with the minimum total distance, which corresponds to the smallest set of errors matching the syndromes. However, the success rate of the MWPM decoder needs to be improved. In this work, we use a GAN decoder to learn the syndrome data of the toric code and, through iterative training, directly generate the result of defect matching elimination.

 Here we consider a depolarizing noise model, in which an input qubit undergoes a completely random Pauli error ($X$, $Y$, or $Z$) with a certain probability $p$, and with probability $1 - p$ the qubit remains unchanged. Mathematically, it can be written as
\begin{equation}{\cal E}\left( \rho  \right) = \left( {1 - p} \right)\rho  + \frac{p}{3}\left( {X\rho X + Y\rho Y + Z\rho Z} \right),\end{equation}	
where $\rho$ is the input qubit state, $p$ is the depolarizing probability, and $X$, $Y$, $Z$ are the Pauli operators. 

Correcting independent X errors on the lattice C and correcting independent Z errors on the dual lattice $C^*$ is  completely equivalent. In this GAN decoder, the focus is on generating the result of defect matching elimination, thus training a decoder model for one type of error is sufficient.

\begin{figure*}
\includegraphics[width=15cm]{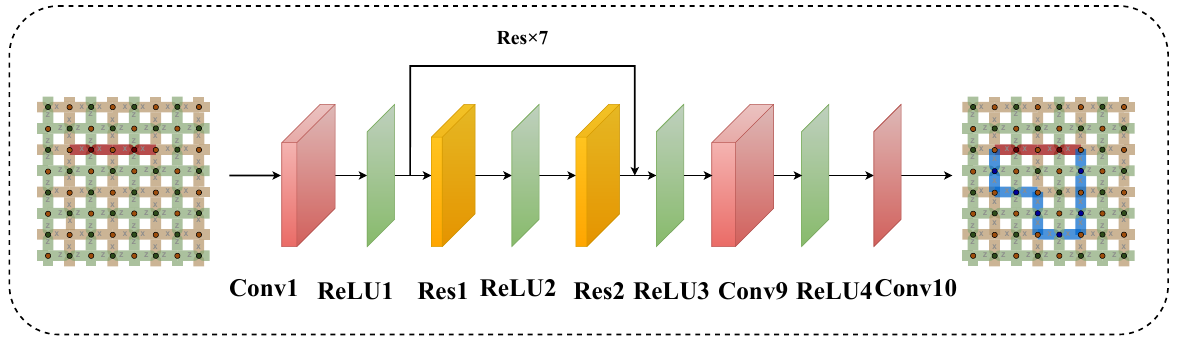}
\caption{\label{f6} The generator structure of the GAN decoder. The red blocks represent convolutional layers, the green blocks represent ReLU activation functions, and the yellow blocks represent residual layers. The parameters of these network layers are listed in table \ref{t1}}
\end{figure*}

\subsection{GAN Decoder}
\subsubsection{Method Framework}

The training framework for this decoder is shown in Figure \ref{f5}. This GAN decoder employs adversarial learning to enable the generator to produce correction paths with trivial loops. The eigenvalues of the toric code forming trivial loops are sent as input to the discriminator in matrix form. 

The discriminator is trained to identify the input as real to update its parameters. The generator is trained with the generated data so that the discriminator identifies it as fake, and the discriminator parameters are updated. Then, the generator is trained to make the discriminator identify the generated data as real, updating the parameters within the generator. By iteratively training the discriminator and the generator in this manner, the process continues until the loss function converges.

\subsubsection{Structure of the generator}

In this GAN decoder, the generator utilizes a convolutional neural network based on residual blocks, as shown in Figure \ref{f6}. Given an input matrix \( X \) and a convolution kernel \( K \), the result of the convolution operation \( Y \) can be represented as:
\begin{equation}
    Y[i, j] = \sum_{m} \sum_{n} X[i+m, j+n] \cdot K[m, n],
\end{equation}
where \( i, j \) denote the position in the output image, and \( m, n \) are the indices of the convolution kernel.  Given an input \( x \), the output of the residual block \( y \) can be expressed as:
\begin{equation}
y = F(x) + x,
\end{equation}
where \( F(x) \) represents the non-linear transformation through the convolutional layers, and activation functions. \( x \) is the input to the residual block, added directly to the output via the shortcut connection.

The eigenvalue matrix of the toric code is concatenated along the channel dimension and is sent as input to the neural network. Using feature maps from various layers, the $n \times n$ matrix is expanded to $128 \times 128$. Here, the number of channels is set to $3$, resulting in an input matrix of $128 \times 128 \times 3$. After an initial convolution layer (Conv1), the input passes through seven residual blocks (Res1-7), followed by two additional convolutional layers (Conv9-10) to extract high-dimensional features. 

The parameters of the network are detailed in table \ref{t1}. To ensure the eigenvalue matrix size remains unchanged, the step is set to $1$. Each convolutional layer is followed by a ReLU activation function \cite{34}
\begin{equation}
\begin{aligned}
\text{ReLU}(x) = 
\begin{cases}
x, & \text{if } x > 0 \\
0, & \text{if } x \leq 0
\end{cases},
\end{aligned}
\end{equation}
where $x$ is the input of the layer.

\begin{table}[h]
\caption{\label{t1}Parameters of the generator}
\tabcolsep 10pt 
\begin{ruledtabular}
\begin{tabular}{ccc}
\textrm{Layer}&
\textrm{Filter size / step}&
\textrm{Output size}\\
\colrule
Conv1&	3×3 / 1	&$128\times128\times64$\\
Res1&	3×3 / 1	&$128\times128\times64$\\
Res2&	3×3 / 1	&$128\times128\times64$\\
Res3&	3×3 / 1	&$128\times128\times64$\\
Res4&	3×3 / 1	&$128\times128\times64$\\
Res5&	3×3 / 1	&$128\times128\times64$\\
Res6&	3×3 / 1	&$128\times128\times64$\\
Res7&	3×3 / 1	&$128\times128\times64$\\
Conv9&	3×3 / 1	&$128\times128\times256$\\
Conv10&	3×3 / 1	&$128\times128\times1$\\
\end{tabular}
\end{ruledtabular}
\end{table}

\subsubsection{Structure of the discriminator}

\begin{figure*}
\includegraphics[width=16cm]{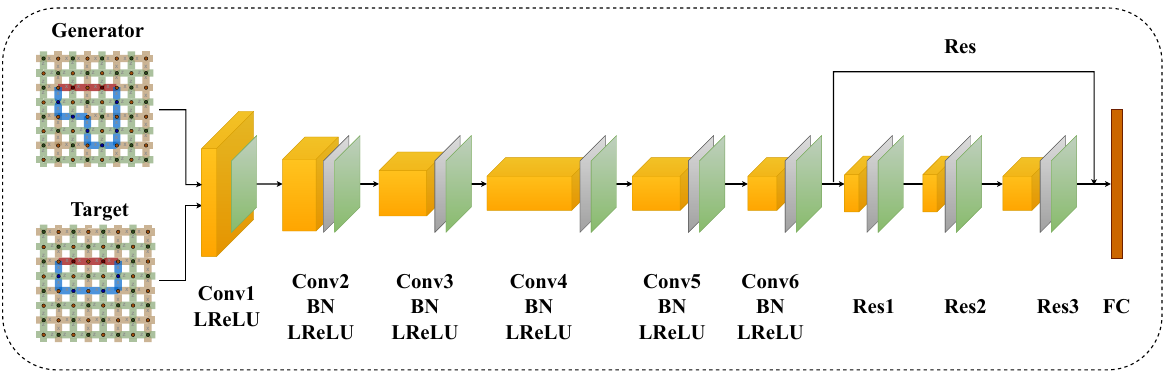}
\caption{\label{f7} The discriminator structure of the GAN decoder. The yellow blocks represent convolutional layers, the green blocks represent LReLU functions, the gray blocks represent BN, and the orange rectangles represent FC layers. The parameters of these network layers are listed in table \ref{t2}}
\end{figure*}

The discriminator structure of the GAN decoder is shown in Figure \ref{f7}. The discriminator uses strided convolutions to extract deep features from both generated and real data (Conv1-6). In the residual blocks, due to changes in data dimensions, a $1 \times 1$ convolution kernel is used (Res1-3), followed by a $3 \times 3$ convolution kernel for feature extraction. Batch Normalization (BN) \cite{35} is applied after each convolution, except for Conv1, to normalize the data. BN normalizes the intermediate activations in each layer during training, which accelerates the training process, improves model stability, and enhances generalization ability.  For each mini-batch input \( x \), the batch normalization process consists of the following steps:

    \textbf{1) Compute Mini-Batch Mean and Variance:}
    \begin{equation}
    \mu_B = \frac{1}{m} \sum_{i=1}^{m} x_i, \quad \sigma_B^2 = \frac{1}{m} \sum_{i=1}^{m} (x_i - \mu_B)^2,
    \end{equation}
    where \( m \) is the number of samples in the mini-batch, and \( x_i \) is the \( i \)-th sample in the mini-batch.

    \textbf{2) Normalize:}
    Normalize each input \( x_i \) by subtracting the mean \( \mu_B \) and dividing by the standard deviation \( \sqrt{\sigma_B^2 + \epsilon} \) (where \( \epsilon \) is a small constant to prevent division by zero):
    \begin{equation}
    \hat{x}_i = \frac{x_i - \mu_B}{\sqrt{\sigma_B^2 + \epsilon}}.
    \end{equation}

    \textbf{3) Scale and Shift:}
    Introduce two learnable parameters, \( \gamma \) (scale parameter) and \( \beta \) (shift parameter), which have the same dimension as the input, allowing the model to retain the representational capacity of the input:
    \begin{equation}
    y_i = \gamma \hat{x}_i + \beta.
    \end{equation}

After BN, the LReLU \cite{36} activation function is used.
\begin{equation}
\begin{aligned}
\text{LReLU}(x) = 
\begin{cases}
x, & \text{if } x > 0 \\
\alpha x, & \text{if } x \leq 0
\end{cases},
\end{aligned}
\end{equation}
where $x$ is the input of the layer, and $\alpha$ is a small constant.

Finally, the network outputs through a fully connected layer (FC). Given an input vector \( x \) of size \( n_{\text{in}} \), the output vector \( y \) of the fully connected layer is computed as:
\begin{equation}
  y = f(Wx + b)  ,
\end{equation}
where \( x \) is the input vector, \( W \) is the weight matrix has a shape of \( (n_{\text{in}}, 1) \),  \( b \) is the bias vector, \( f \) is the sigmoid activation function,
\begin{equation}
     \sigma(x) = \frac{1}{1 + e^{-x}}.
\end{equation}
The parameters of the network are detailed in table \ref{t2}.

\begin{table}[h]
\caption{\label{t2}Parameters of discriminator}
\tabcolsep 10pt 
\begin{ruledtabular}
\begin{tabular}{ccc}
\textrm{Layer}&
\textrm{Filter size / step}&
\textrm{Output size}\\
\colrule
Conv1&	3×3 / 1&	128×128×64\\
Conv2&	3×3 / 2&	64×64×128\\
Conv3&	3×3 / 2&	32×32×256\\
Conv4&	3×3 / 2&	16×16×512\\
Conv5&	3×3 / 1&	16×16×256\\
Conv6&	1×1 / 1&	16×16×128\\
Res1&	1×1 / 1&	16×16×64\\
Res2&	3×3 / 1&	16×16×64\\
Res3&	3×3 / 1&	16×16×128\\
FC	&   -	   &     1    \\
\end{tabular}
\end{ruledtabular}
\end{table}

\begin{figure*}
	\includegraphics[width=14.5cm]{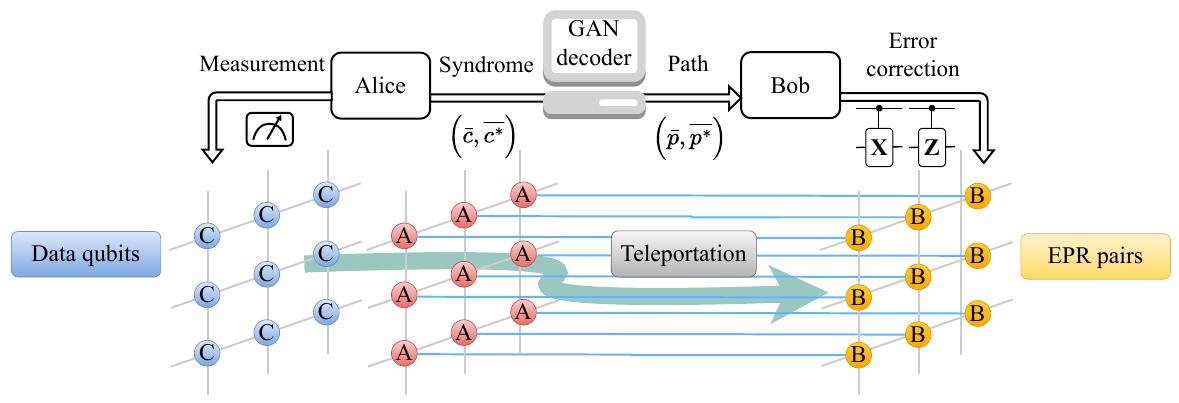}
	\caption{\label{f12}The toric code optimizes quantum teleportation using a GAN decoder. Data qubits are teleported to Bob. Alice measures the toric code and sends the syndrome $\left( \bar{c}, \overline{c^*} \right)$ to the GAN decoder. The decoder outputs the error correction path, and Bob applies the corresponding error correction operations to the qubits on the path.}
\end{figure*}

\subsubsection{Loss functions of the GAN decoder}
In deep learning, loss functions are used to measure the discrepancy between generated data and the target data. Let the generator be represented as $G_\theta$ and the discriminator as $D_\phi$, where $\theta$ and $\phi$ are the network parameters for $G_\theta$ and $D_\phi$, respectively. The optimization of GANs can be defined in various ways. In this work, we consider the non-saturation loss \cite{37}. Sample $m$ data ${g^i}\left( \theta  \right)$ from the generator and $m$ random data from the dataset, the loss of the generator is
\begin{equation}
\label{e5}
\begin{aligned}
{L_G}\left( {\phi ,\theta } \right) =  - \frac{1}{m}\sum\limits_{i = 1}^m {\left[ {\log {D_\phi }\left( {{g^i}\left( \theta  \right)} \right)} \right]} .
\end{aligned}
\end{equation}
The loss of the discriminator is
\begin{equation}
\label{e6}
\begin{aligned}
{L_D}\left( {\phi ,\theta } \right) = \frac{1}{m}\sum\limits_{i = 1}^m {\left[ {\log {D_\phi }\left( {{{\rm{y}}^i}} \right) + \log \left( {1 - {D_\phi }\left( {{g^i}\left( \theta  \right)} \right)} \right)} \right]} .
\end{aligned}
\end{equation}
The process of training GANs is equivalent to finding the Nash equilibrium 
\begin{equation}
\begin{aligned}
\mathop {\max }\limits_\theta  {L_G}\left( {\phi ,\theta } \right),\mathop {\max }\limits_\phi  {L_D}\left( {\phi ,\theta } \right).
\end{aligned}
\end{equation}
Alternate training is used to optimize $\theta$ of the generator and $\phi$ of the discriminator.

\subsection{Optimization of teleportation}

Figure \ref{f12} illustrates how the toric code with a GAN decoder optimizes quantum teleportation.
When Alice performs quantum teleportation of toric code data qubits to Bob, noise may introduce errors in the data qubits. In such cases, the GAN decoder provides Bob with an error correction strategy based on the syndromes. Bob then applies this strategy to correct the errors, thereby enhancing the fidelity of quantum teleportation under noisy conditions.

This method involves three main processes. First, the initialization process prepares the quantum state prior to teleportation. Second, the teleportation process transmits the data qubits. Third, errors that occur during transmission are detected and corrected. The following provides a detailed description of these three processes.

\subsubsection{Initialization process}

Before Alice and Bob execute teleportation, $d^2$ EPR pairs are prepared
\begin{equation}
\begin{aligned}
\left| \Psi  \right\rangle _{{\rm{AB}}}^i = \frac{1}{{\sqrt 2 }}\left( {\left| {00} \right\rangle _{ab}^i + \left| {11} \right\rangle _{ab}^i} \right),
\end{aligned}
\end{equation}
where $d$ is the code distance of the quantum toric code, $i = 1,2, \cdots ,{d^2}$. Alice holds set A of particles, and Bob holds set B of particles. The states we aim to teleport are the data qubits of toric code, consisting of $d^2$ physical qubits, denoted as $ \left| \phi  \right\rangle$. Each data qubit is teleported using a pair of EPR states.

%

\subsubsection{Teleportation process}
Alice teleports the state $\left| {\phi } \right\rangle $ to Bob by three steps:

1) Alice sets the particles in  $\left| { \phi } \right\rangle $ as control qubits, and the corresponding particles of set A as target qubits, execute CNOT gates.

2) Alice applies H gates to the particles in set A and measures both the particles of set A and the particles of $\left| \phi  \right\rangle $, obtaining $d^2$ pairs of outcomes. Every pair of outcome corresponds to one of the results in $\left\{ {00,01,10,11} \right\}$ with a probability of $1/4$. The outcomes are sent to Bob via a classical communication channel.

3) After the measurement, the information of  $\left| \phi  \right\rangle $ is teleported to the particles in set B. According to the outcomes, Bob performs the associated recovery gates on the particles in set B, as shown in table \ref{t3}.

\begin{table}[h]
\caption{\label{t3} The outcomes of teleportation and the recovery gates}
\tabcolsep 6pt 
\begin{ruledtabular}
\begin{tabular}{cccc}
\textrm{Particles in $\left| \psi  \right\rangle $}&
\textrm{Particles in set A}&
\textrm{Outcomes}&
\textrm{Gates}\\
\colrule
$\left| 0 \right\rangle^i $&$\left| 0 \right\rangle _a^i$	&00&I\\
$\left| 0 \right\rangle^i $&$\left| 1 \right\rangle _a^i$	&01&X\\
$\left| 1 \right\rangle^i $&$\left| 0 \right\rangle _a^i$	&10&Z\\
$\left| 1 \right\rangle^i $&$\left| 1 \right\rangle _a^i$	&11&Y\\
\end{tabular}
\end{ruledtabular}
\end{table}

After Bob performs the recovery gates on the particles in set B, the states ${\left| {\tilde \phi } \right\rangle  }$ is obtained. However, due to noise and imperfect quantum operations, ${\left| {\tilde \phi } \right\rangle }$ may not be equal to $\left| { \phi } \right\rangle $.

\subsubsection{Error correction process}
In this process, Bob detects the received ${\left| {\tilde \phi } \right\rangle }$ and checks whether errors occur. Any data qubits of ${\left| {\tilde \phi } \right\rangle }$ with stabilizers $Z_v$ and $X_f$ has
\begin{equation}
\begin{aligned}
Z_v^1Z_v^2Z_v^3Z_v^4{\left| {{\varphi _i}} \right\rangle }  = \hat Z{\left| {{\varphi _i}} \right\rangle } ,
\end{aligned}
\end{equation}
\begin{equation}
\begin{aligned}
X_f^1X_f^2X_f^3X_f^4{\left| {{\varphi _i}} \right\rangle }  = \hat X{\left| {{\varphi _i}} \right\rangle } ,
\end{aligned}
\end{equation}

If no error occurs, due to the commutation between the stabilizers of the same kind, the eigenvalues of  $Z_v$ and $X_f$ are stable, $\hat Z= +1$, $\hat X = +1$. On the contrary, if errors occur, due to the anti-commutation between errors and stabilizers, the eigenvalues of adjacent stabilizers change from $+1$ to $-1$, $\hat Z= - 1$, $\hat X= - 1$.

Therefore, by observing the adjacent stabilizers of data qubits, Bob obtains two sets of eigenvalue measurement results $\left( {\bar c,\overline {{c^ * }} } \right)$ corresponding to X errors and Z errors. $\left( {\bar c,\overline {{c^ * }} } \right)$ are the input of the generator in the GAN decoder, and the output $\left( {\bar p,\overline {{p^ * }} } \right)$ of the generator is the decoding result.

\section{Evaluation of simulation experiments}\label{sec:level4} 

\subsection{Model training}
\begin{figure}
	\includegraphics[width=9cm]{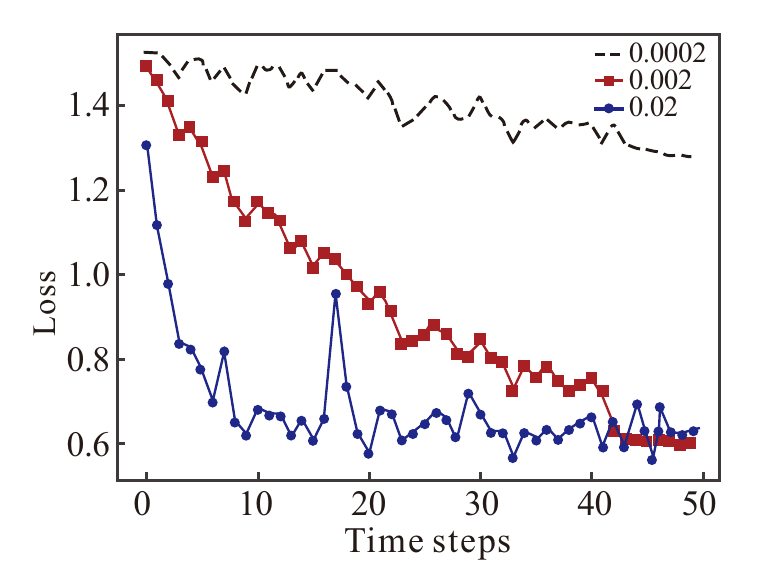}
	\caption{\label{frate}Training performance at different learning rates. The black dotted line shows the loss function at a learning rate of 0.0002; the red line at 0.002; and the blue line at 0.02.}
\end{figure}
\begin{figure*}
	\includegraphics[width=17cm]{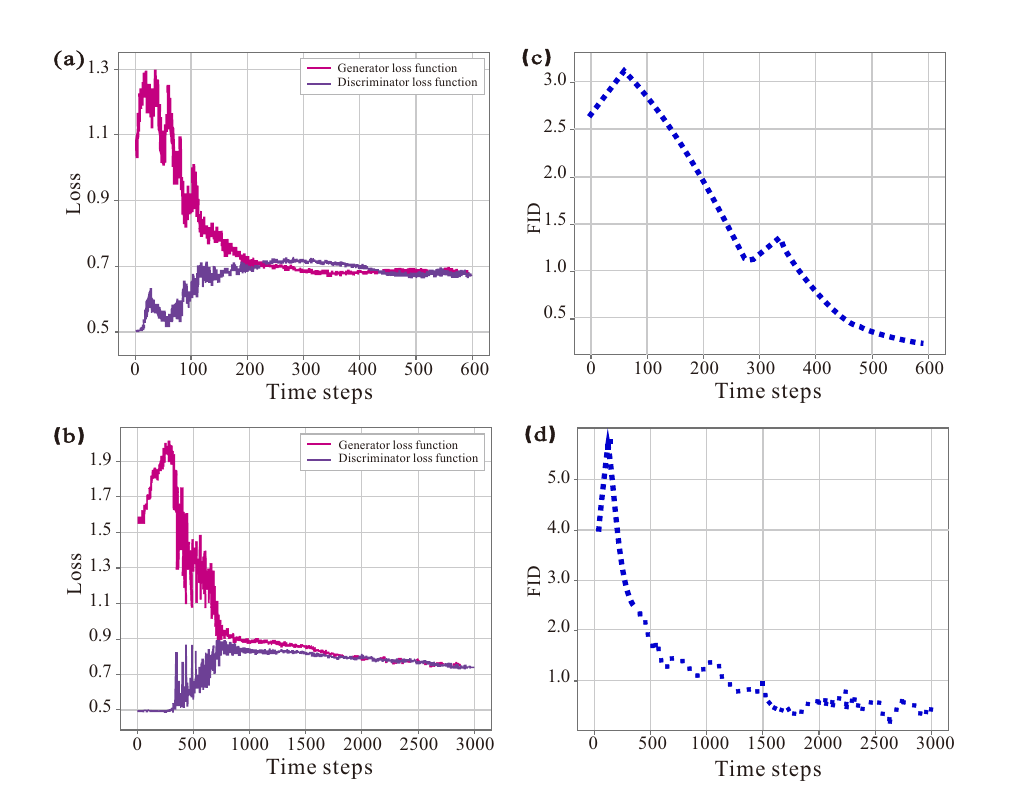}
	\caption{\label{f8}Training results. (a) Loss function progression for the $d=3$ GAN decoder. (b) FID progression for the the $d=3$ GAN decoder. (c) Loss function progression for the $d=5$ GAN decoder. (d) FID progression for the the $d=5$ GAN decoder.}
\end{figure*}

The experimental environment is configured with an Intel Xeon E5-2640V4 processor at 2.4 GHz, 32 GB of cache, and a GTX1080Ti GPU. The programming language and model building platform used are Python and Anaconda. The training framework is based on TensorFlow, Qiskit, and Keras.

Before training the decoder, the datasets are generated as follows. First, the topology of the toric code is defined, including the lattice with code distance $d$, physical qubits, and stabilizers. Second, an independent noise model is applied with varying error rates to randomly inject errors onto the qubits. The syndrome is obtained by measuring the stabilizers. Next, the corresponding path that forms a trivial cycle with the syndrome is calculated. The training data are labeled as (syndrome, path) for decoder training.

The sample preprocessing includes selecting small batch data, data augmentation operations, and standardization processing. Data augmentation mainly includes left-right flipping and up-down flipping, with the random function generating arbitrary floating-point numbers between 0 and 1. In this simulation experiment, the flipping probability is set to 0.25. Additionally, batch size is set to 4.

The learning rate is a hyperparameter that measures the speed of gradient descent in the network. An appropriate learning rate enables the network to converge quickly, reducing training time. However, if the learning rate is too high, oscillations may occur, preventing the network from converging. Conversely, if the learning rate is too low, the convergence speed of the network during iterations will be slow, resulting in a longer training time. We test different learning rates to observe their impact on the loss function.

Figure \ref{frate} shows the training performance at different learning rates. When the learning rate is set to 0.0002, the loss value decreases very slowly, requiring more time to achieve the desired result. At a learning rate of 0.02, the loss value decreases rapidly, but oscillations occur, which affect the network's convergence. When the learning rate is 0.002, the convergence is better: the loss value decreases quickly and remains stable. Therefore, we set the learning rate to 0.002 for training.

Next, we employ the Adam optimization algorithm to train the network parameters. Adam uses both the first and second moment estimates during training, making it a robust optimization technique for non-smooth objective functions and noisy gradients \cite{38}. According to Eqs. (\ref{e5}) and (\ref{e6}), we use the loss function to show whether our method converged. We calculate the distance between target data and generate data in the feature space using the Frechet Inception Distance (FID) \cite{heusel2017gans}
\begin{equation}
{\rm{FID}}\left( {{t},{g}} \right) = \left\| {{\mu _t},{\mu _g}} \right\| + \mathrm{Tr}\left( {{C_t} + {C_g} - 2{{\left( {{C_t}{C_g}} \right)}^{1/2}}} \right),
\end{equation}
where ${\mu _t}$ and ${\mu _g}$ are the averages of the feature vectors of the target data and the generated data, respectively. ${C_t}$ and ${C_g}$ are their respective covariance matrices.  $\mathrm{Tr}\left[  \cdot  \right]$ denotes the trace of a matrix. $\left\|  \cdot \right\|$ denotes the $L_2$ norm of a vector.

Figure \ref{f8} shows the training results of the GAN decoder. Figures 10a and 10b display the convergence of the generator and discriminator losses. The GAN decoder models for the quantum toric codes with $d=3$ and $d=5$ were considered sufficiently trained after 500 and 1800 training iterations, respectively. After averaging over samples, the resulting loss values typically fall within the range $(0, 2)$. In our training, the final loss value converges to approximately 0.7, which reflects a balanced optimization between the generator and discriminator networks. This value indicates that the model has learned to generate outputs that are reasonably close to the target distribution while maintaining sufficient discriminator sensitivity. 
Additionally, Figures 10c and 10d illustrate the changes in FID with iterations, demonstrating how the distribution of the generated data evolves toward the distribution of the target data. After 500 and 1800 training iterations for $d=3$ and $d=5$ respectively, the FID approaches convergence, indicating that the distribution of the generated samples is close to that of the training samples. 

Specifically, the generator of the GAN decoder contains approximately 409,153 trainable parameters, while the discriminator includes about 2,914,433 parameters. Table \ref{t5}  shows the average decoding time per sample and the average training time per iteration for GAN decoders with code distances d=3 and d=5. As the code distance increases, both the training time and the decoding time increase. Although the number of trainable parameters remains fixed, larger code distances result in higher computational costs and greater memory usage.

As the code distance increases, the size of the input syndrome and the GAN networks grows, leading to higher computational and memory requirements. In practice, the training time scales roughly quadratically to cubically with code distance, depending on the network architecture and dataset size. However, once trained, the model can be reused for multiple decoding tasks, and incremental fine-tuning can reduce retraining costs for slightly larger distances.

During inference, the GAN decoder performs a single forward pass through the generator network for each syndrome. Therefore, the decoding time scales approximately linearly with the number of qubits (proportional to the square of the code distance for surface codes).

\begin{table}[h]
\caption{\label{t5}The decoding time and training time of the proposed GAN decoder.}
\begin{tabular}{p{2.2cm}<{\centering}p{3cm}<{\centering}p{2.8cm}<{\centering}}
\toprule
\textrm{Code distance}&
\textrm{Avg. decoding time / sample}&
\textrm{Avg. training time / iteration}\\
\colrule
3&	5.13 msec &	 127.83 s\\
5&  10.79 msec & 326.56 s\\
\toprule
\end{tabular}
\end{table}

\subsection{Result evaluation}
\begin{figure}
\includegraphics[width=8cm]{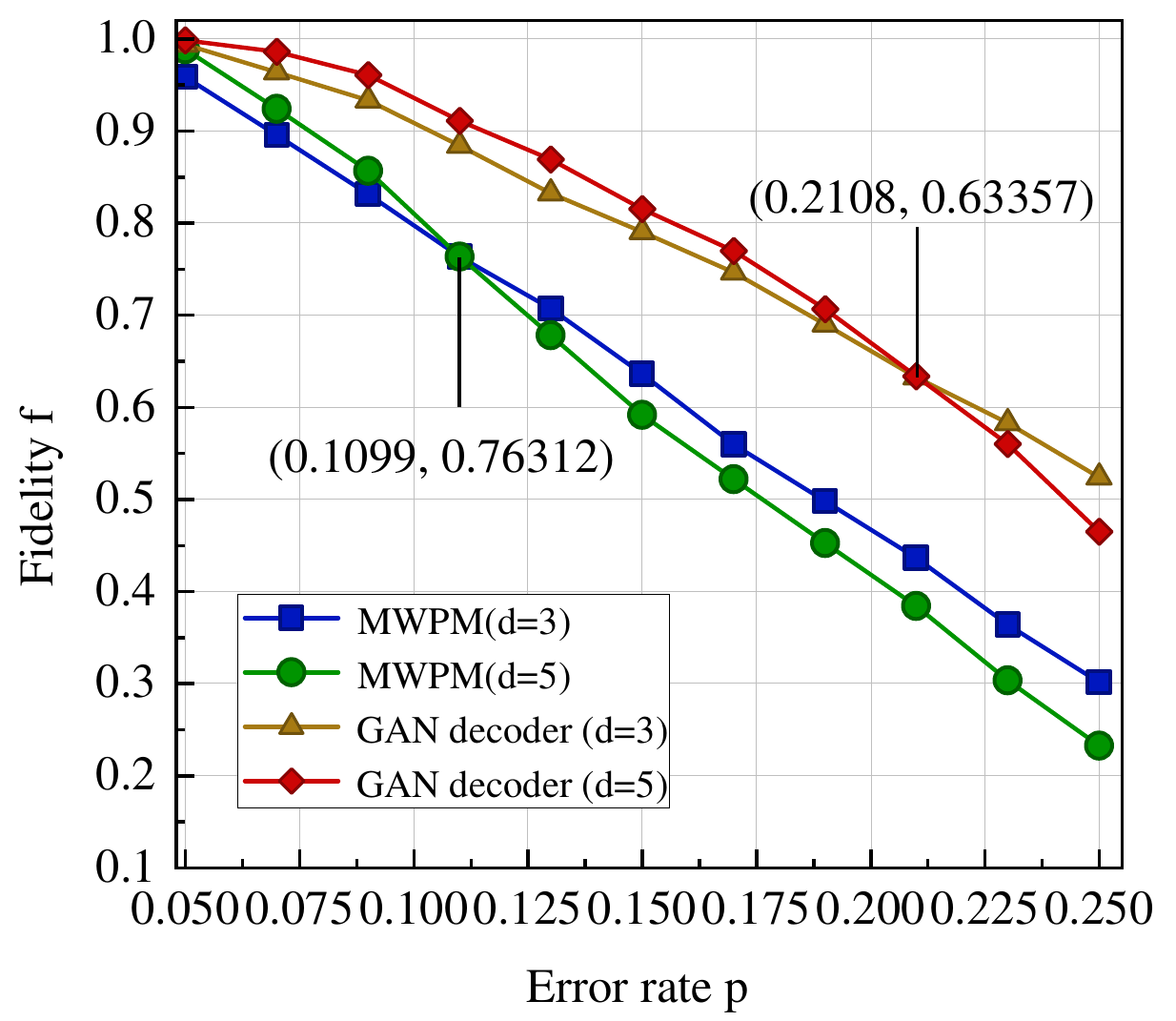}
\caption{\label{f9} Evaluation of the model. Logical fidelity $f$ of GAN decoder versus bit flip error rate $p$ for system size $d = 3$, $d = 5$ and compared to corresponding results using MWPM.}
\end{figure}

\begin{figure}
\includegraphics[width=8cm]{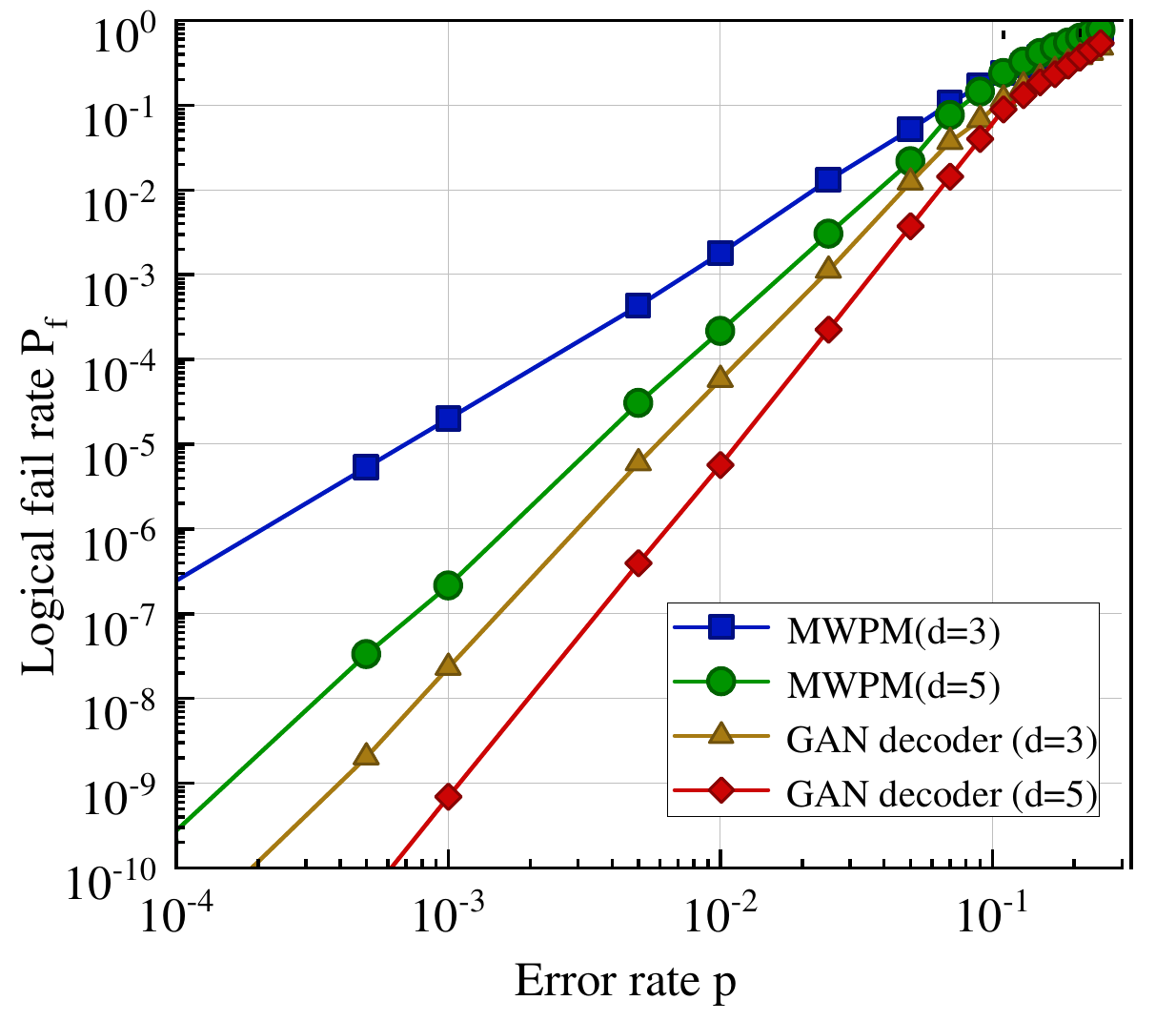}
\caption{\label{f9a} 
 Logical fail rate $P_f=1-f$ of GAN decoder versus bit flip error rate $p$ for system size $d = 3$, $d = 5$ and compared to corresponding results using MWPM.}
\end{figure}

By training the GAN model, we achieved a logical fidelity of 99.895\% at $p=0.05$ for the toric code decoder with code distance $d=5$. We tested the error correction capabilities of the trained GAN model and the local MWPM model on toric code lattices with code distances $d=3$ and $d=5$.  Considering that the encoding scheme of toric codes is not unique, all comparative experiments in our study use the same encoding scheme to ensure fairness and consistency. 
The test results are presented in Figures \ref{f9} and \ref{f9a}, illustrating the optimization performance and the logarithmic behavior of the GAN decoder in comparison to the MWPM decoder. As the error rate increase, the error correction success rate significantly decreases. 

The impact of code distance on logical fidelity can be characterized by the pseudo-threshold. In this work, the pseudo-threshold of the quantum toric code decoder is estimated based on the intersection point of the logical error rate curves for code distances d=3 and d=5. This crossing point provides only an approximate finite-size estimate for the depolarizing noise model and higher than the theoretical threshold of the toric code, which is known to be below 11\%\cite{33}. Below the threshold, the logical error rate decreases as the code distance grows. Conversely, above the threshold, the logical error rate either stops improving or even increases, rendering error correction ineffective.

Under the same conditions, the pseudo-threshold for the local MWPM model is approximately $P=0.1099$, which is close to that of the RL decoder \cite{10b} and the RNN decoder ($P=0.1363$) \cite{10a}, while our GAN-based decoder achieves a significantly higher pseudo-threshold of $P=0.2108$.
Overall, in terms of logical fidelity performance, the GAN model outperforms these existing decoders for decoding toric codes with $d=3$ and $d=5$.


Additionally, we verify the optimized quantum teleportation method using topological codes with the GAN decoder. The quantum operations in the teleportation algorithm are set to be noisy. We test the fidelity performance of qubits, based on topological code error correction, in quantum circuits under different noise conditions. As shown in Table \ref{t4}, in the original quantum teleportation fidelity simulation experiment, the fidelity of the quantum state decreases more rapidly with increasing noise. In contrast, in the simulation experiment using toric codes with the GAN decoder, the decrease in quantum state fidelity is alleviated as noise increases.

\begin{table*}[htbp] 
	\caption{ Simulation Results of Quantum Teleportation for the $d=5$ GAN Decoder}
	\label{t4}
	\centering
	\begin{tabular}{p{2cm}<{\centering}p{2cm}<{\centering}p{0.8cm}<{\centering}p{0.8cm}<{\centering}p{0.8cm}<{\centering}p{0.8cm}<{\centering}p{0.8cm}<{\centering}p{0.8cm}<{\centering}p{2.3cm}<{\centering}} 
		\toprule
		\multirow{2}{1.7cm}{Error rate ($\times10^{-3} $)} & \multirow{2}{2.1cm}{Original fidelity (\%)} & \multicolumn{6}{c}{Number of failures at d=5 ($n_\mathrm{shoot}=1024$)} & \multirow{2}{2.2cm}{Optimizing fidelity (\%)} \\ \cline{3-8}
		&                                      & 1       & 2       & 3       & 4       & 5       & Ave       &                                        \\ \hline
		5                           & 99.52                                & 0       & 0       & 0       & 0       & 0       & 0         & 99.99                                    \\ 
		10                          & 99.11                                & 1       & 0       & 0       & 0       & 0       & 0.2       & 99.98                               \\
		15                          & 98.63                                & 1       & 1       & 3       & 4       & 3       & 2.4       & 99.77                               \\
		20                          & 97.97                                & 6       & 2       & 5       & 5       & 3       & 4.2       & 99.58                               \\
		25                          & 97.53                                & 9       & 6       & 6       & 3       & 7       & 6.2       & 99.39                               \\
		30                          & 96.98                                & 11      & 8       & 2       & 13      & 9       & 8.6       & 99.16                               \\
		35                          & 96.52                                & 11      & 16      & 18      & 5       & 21      & 14.2      & 98.62                               \\
		40                          & 95.97                                & 22      & 23      & 20      & 21      & 18      & 20.8      & 97.97                               \\
		45                          & 95.46                                & 26      & 27      & 22      & 24      & 28      & 25.4      & 97.52                               \\
		50                          & 94.95                                & 29      & 31      & 33      & 28      & 30      & 30.2      & 97.05                               \\
		55                          & 94.52                                & 47      & 41      & 39      & 31      & 26      & 36.8      & 96.40                               \\
		60                          & 93.95                                & 34      & 45      & 53      & 43      & 44      & 46        & 95.52                               \\
		65                          & 93.53                                & 55      & 61      & 54      & 47      & 51      & 53.6      & 94.76                               \\
		70                          & 93.14                                & 72      & 78      & 67      & 53      & 61      & 66.2      & 93.54                               \\
		75                          & 92.66                                & 79      & 74      & 63      & 71      & 74      & 72.2      & 92.95                               \\ 
		80                          & 92.23                                & 61      & 76      & 94      & 86      & 95      & 82        & 91.99                               \\ 	
\toprule
	\end{tabular} 
\end{table*}

\begin{figure}
	\includegraphics[width=9cm]{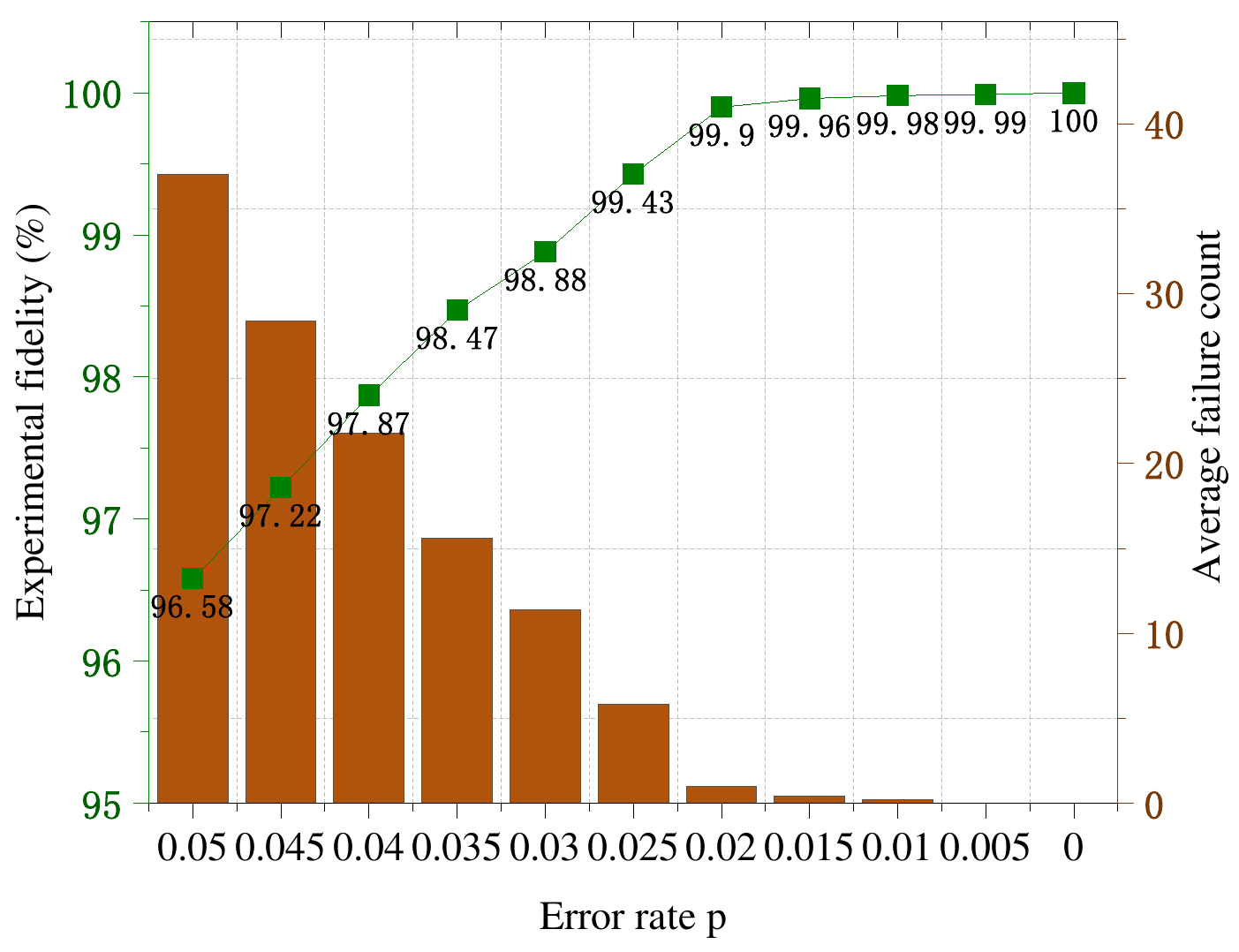}
	\caption{\label{f10} The trend of quantum state fidelity and the number of uncorrectable errors versus the physical error rate $p$}. The red bar chart shows the number of uncorrectable errors when $n_\mathrm{shoot} = 1024$, and the green curve represents the fidelity performance in the simulation experiment.
\end{figure}

In this simulation experiment, when the error rate $p = 5 \times {10^{ - 3}}$, the toric code with the GAN decoder provides the fidelity performance better than 99.99\%.
Quantum states that fail error correction will be detected and rejected. Figure \ref{f10} shows the fidelity of the quantum state and the number of failed error correction measurements as the error rate of the noise model changes, with $n_\mathrm{shoot}=1024$.

\begin{figure}
	\includegraphics[width=8.5cm]{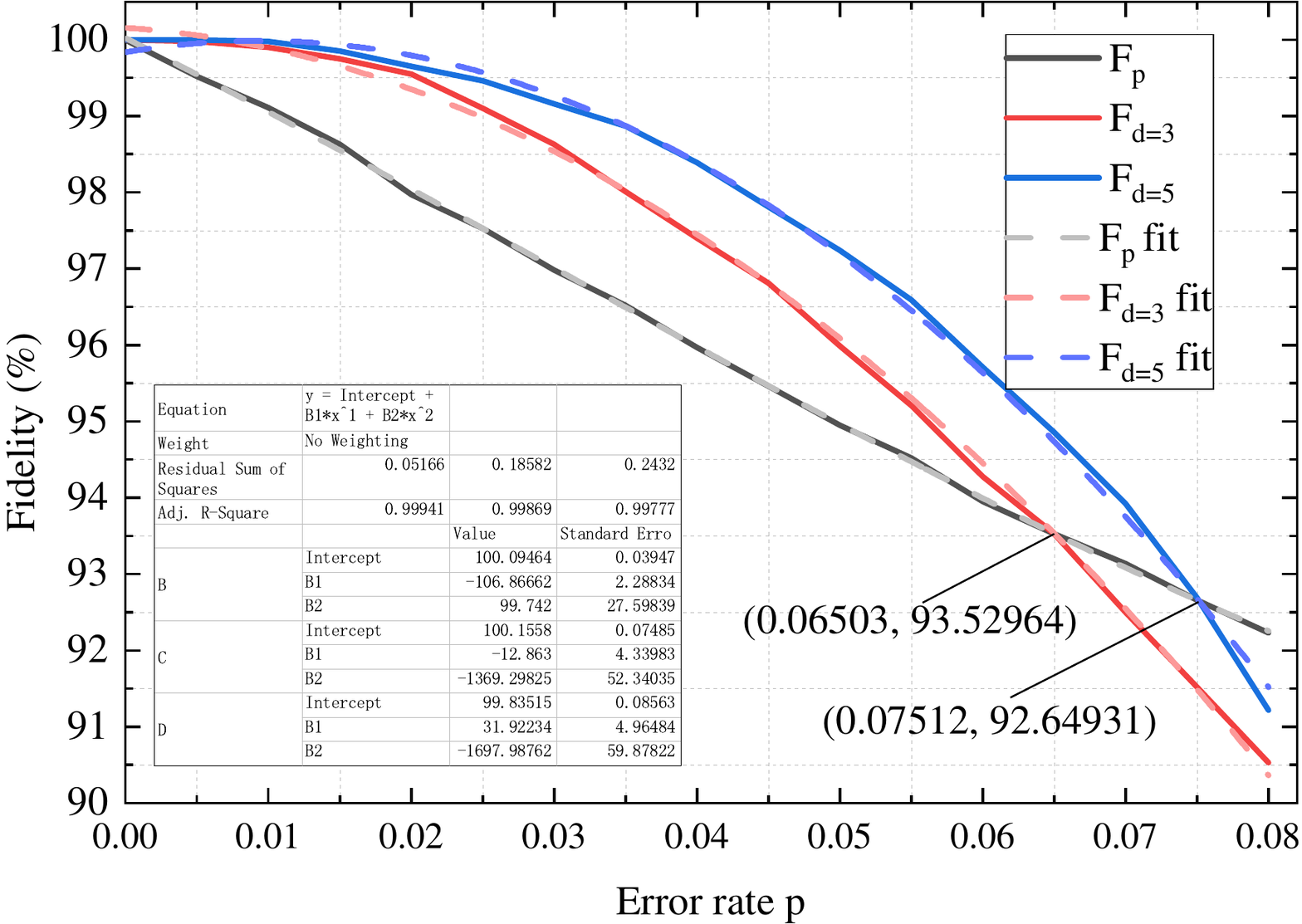}
	\caption{\label{f11} Comparison of the fidelity variation curves between original quantum teleportation and GAN decoder optimized quantum teleportation. The black solid line represents the fidelity performance of the original teleportation, the red solid line represents the fidelity performance of the optimized model with $d=3$, and the blue solid line represents the fidelity performance of the optimized model with $d=5$. The black, red, and blue dashed lines are the fitting curves used for data analysis.}
\end{figure}

\begin{figure}
	\includegraphics[width=8.5cm]{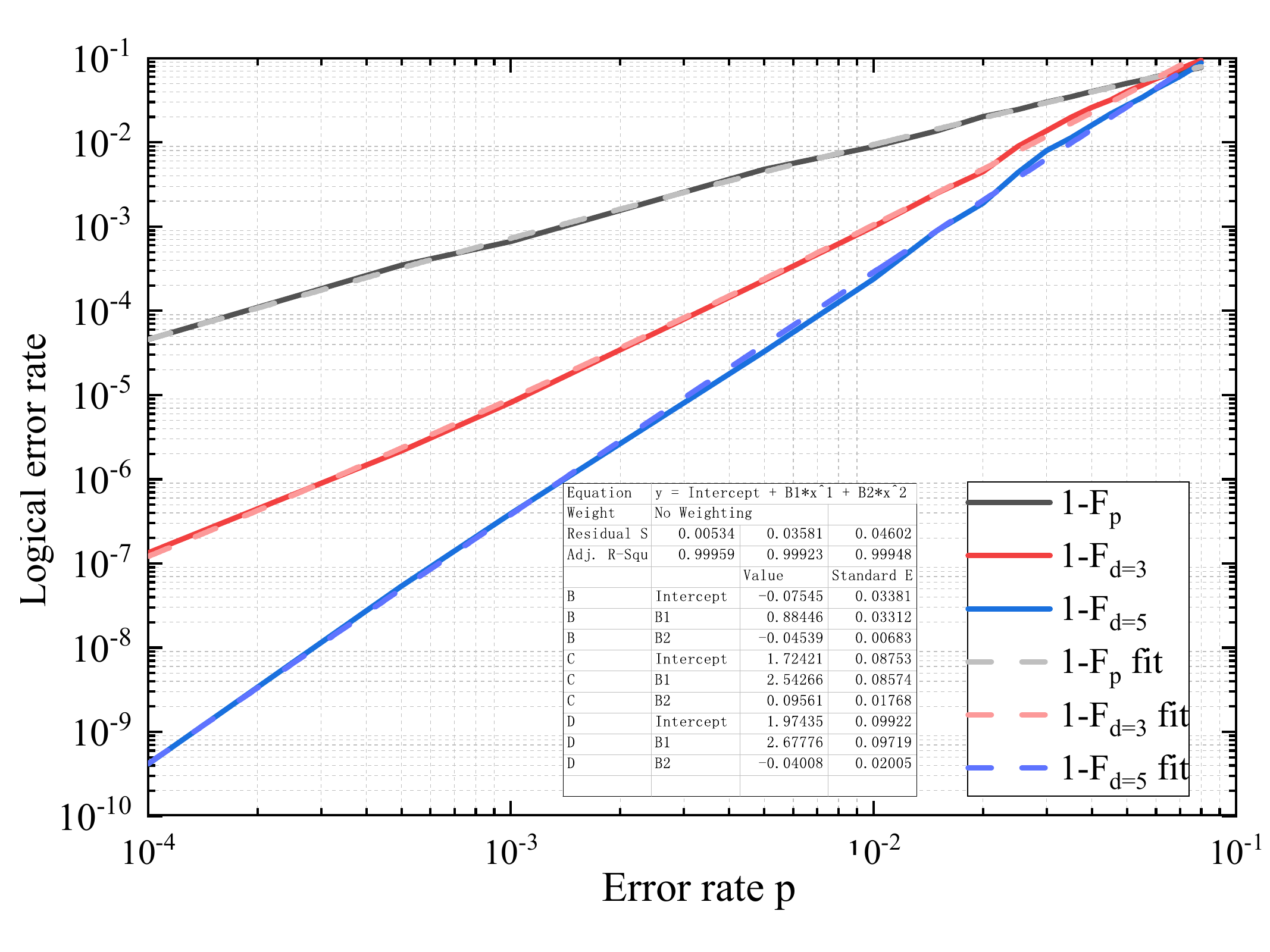}
	\caption{\label{f11a} Comparison of the logical error rate curves between original quantum teleportation and GAN decoder optimized quantum teleportation. The black solid line represents the logical error rate performance $1-{F_p}$ of the original teleportation, the red solid line represents the logical error rate performance $1-{F_{d=3}}$  of the optimized model with $d=3$, and the blue solid line represents  the logical error rate performance $1-{F_{d=5}}$ of the optimized model with $d=5$. The black, red, and blue dashed lines are the fitting curves used for data analysis.}
\end{figure}

Figures \ref{f11} and \ref{f11a} compare the optimization of the teleportation method using topological codes with the GAN decoder and show the logarithmic behavior of the solution. The figure indicates that in the depolarizing noise simulation experiment, for the system with $d=3$ toric code, the fidelity threshold of the model is approximately 0.06503; for the system with $d=5$ toric code, the fidelity threshold is approximately 0.07512. This implies that the model can ensure effective fidelity improvement in quantum networks with error rates below the threshold in the noise model system.

\section{Conclusion}\label{sec:level5} 

In this work, we have employed a GAN algorithm to train and obtain a suited quantum toric code decoder. For toric code lattices of scale $d=3$ and $d=5$, the error correction success rates of this model reaches 99.739\% and 99.895\%, respectively, at an error rate $p=0.05$. Under the same conditions, our model demonstrates superior fidelity compared to the MWPM decoder. Moreover, it achieves a pseudo-threshold of $p=0.2108$, which is substantially higher than that of the MWPM decoder (0.1099).

In the simulation experiment of optimizing the teleportation protocol, we have observed an effective fidelity enhancement within a noise range where $p<0.07512$ compared to the original teleportation protocol. In general, the GAN decoder shows potential for decoding various types of topological codes, paving the way to broader applications. 

Regarding scalability, the proposed GAN-based decoder can, in principle, be extended to larger code distances by retraining or fine-tuning the model on appropriately generated datasets. The generator–discriminator structure enables parallel inference, which may help mitigate the increase in decoding latency as the system size grows. However, scaling to very large distances will inevitably increase the input dimensionality and model parameters, which could lead to higher training costs and memory consumption. To address this, future work may explore hierarchical GAN architectures or dimensionality reduction techniques to preserve decoding performance while maintaining computational efficiency. Additionally, exploring various GAN architectures, such as Quantum Generative Adversarial Networks (QGANs) \cite{39,40,41}, CycleGAN \cite{42}, Self-Attention GAN (SAGAN) \cite{43}, may improve the efficiency and accuracy of decoders, thereby achieving higher fidelity in a broader range of quantum computing applications.

The current analysis does not explicitly consider measurement errors. However, our proposed GAN-based decoder can be adapted to such scenarios by training on syndrome datasets generated under noise models that include measurement errors. Also, non-homogeneous noise can be taken into account. The performance of the GAN decoder under these conditions will be addressed in a future work to further evaluate its capabilities.

Moreover, applying GAN decoders to the decoding tasks of different error-correcting codes on actual quantum computing hardware is a promising direction. This could lay the groundwork for developing more reliable quantum communication and computation systems in noisy quantum environments. As quantum technology continues to advance \cite{Alberts2021,DeMicheli2022,Bluvstein2024,Monika2024}, GAN-based decoders are expected to become essential tools for efficient quantum error correction, propelling quantum computing towards practical implementation. Therefore, our results supply useful insights towards these further developments \cite{QECNature2024}.


\begin{acknowledgments}
The authors acknowledge the Natural Science Foundation of Shandong Province of China (ZR2021ZD19). R.L.F. acknowledges support by MUR (Ministero dell’Università e della Ricerca) through the following projects: PNRR Project ICON-Q – Partenariato Esteso NQSTI – PE00000023 – Spoke 2 – CUP: J13C22000680006, PNRR Project QUANTIP – Partenariato Esteso NQSTI – PE00000023 – Spoke 9 – CUP: E63C22002180006, PNRR Project AQuSDIT – Partenariato Esteso SERICS – PE00000014 – Spoke 5 – CUP: H73C22000880001, PNRR Project PRISM – Partenariato Esteso RESTART – PE00000001 – Spoke 4 – CUP: C79J24000190004. Besides, Program of China Scholarship Council (Grant NO. 202306330060) are gratefully acknowledged.
\end{acknowledgments}



%

\end{document}